\documentclass[conference]{IEEEtran}
\IEEEoverridecommandlockouts
\usepackage{amsmath,amssymb,amsfonts}
\usepackage{algorithmic}
\usepackage{graphicx}
\usepackage{textcomp}
\usepackage[table,xcdraw]{xcolor}
\usepackage{balance}
\usepackage{soul}
\usepackage{booktabs}
\usepackage{multirow}
\usepackage{float}
\usepackage{url}
\usepackage{tcolorbox}
\usepackage{tabularx} 
\usepackage[noadjust]{cite}
\usepackage[hidelinks,bookmarks=true,pagebackref=true]{hyperref}
\usepackage{xurl}
\usepackage{enumitem}
\setlist[itemize,enumerate]{noitemsep, topsep=0pt, leftmargin=1.0em}
\usepackage[utf8]{inputenc}
\usepackage{newunicodechar,graphicx}

\usepackage[switch]{lineno}
\usepackage{times}
\usepackage[none]{hyphenat}
\usepackage{latexsym}
\usepackage{indentfirst}
\graphicspath{{images/}}
\usepackage{amsmath}
\usepackage{url} 
\usepackage{algorithmic}
\usepackage{xcolor,colortbl}
\usepackage{listings}
\definecolor{dkgreen}{rgb}{0,0.6,0}
\definecolor{gray}{rgb}{0.5,0.5,0.5}
\definecolor{mauve}{rgb}{0.58,0,0.82}
\colorlet{punct}{red!60!black}
\definecolor{background}{HTML}{EEEEEE}
\definecolor{delim}{RGB}{20,105,176}
\colorlet{numb}{magenta!60!black}

\definecolor{stringcolor}{RGB}{163,21,21}
\lstdefinelanguage{json}{
    basicstyle=\normalfont\ttfamily,
    numbers=none,
    numberstyle=\scriptsize,
    stepnumber=1,
    numbersep=8pt,
    showstringspaces=false,
    breaklines=true,
    frame=none,
    literate=
     *{:}{{{\color{punct}{:}}}}{1}
      {,}{{{\color{punct}{,}}}}{1}
      {\{}{{{\color{delim}{\{}}}}{1}
      {\}}{{{\color{delim}{\}}}}}{1}
      {[}{{{\color{delim}{[}}}}{1}
      {]}{{{\color{delim}{]}}}}{1},
      morestring=[b]"
}
\definecolor{Gray}{gray}{0.9}
\definecolor{White}{gray}{1.0}

\usepackage{xspace}
\newcommand{\installID}{\texttt{InstallID}\xspace}
\newcommand{\projectID}{\texttt{ProjectID}\xspace}
\newcommand{\metaComment}{\texttt{metaComment}\xspace}
\newcommand{\infectionStack}{\texttt{InfectionStack}\xspace}
\newcommand{\ToolName}{PasteTrace\xspace}

\begin{document}

\title{\ToolName: A Single Source Plagiarism Detection Tool For Introductory Programming Courses}

\author{
\IEEEauthorblockN{Jesse McDonald}
\IEEEauthorblockA{
Information and Computer Sciences\\
University of Hawai`i at M\=anoa\\
Honolulu, Hawai`i\\
Email: jamcd [at] hawaii.edu
}
\and
\IEEEauthorblockN{Scott Robertson}
\IEEEauthorblockA{
Information and Computer Sciences\\
University of Hawai`i at M\=anoa\\
Honolulu, Hawai`i\\
Email: scottpr [at] hawaii.edu
}
\and
\IEEEauthorblockN{Anthony Peruma}
\IEEEauthorblockA{
Information and Computer Sciences\\
University of Hawai`i at M\=anoa\\
Honolulu, Hawai`i\\
Email: peruma [at] hawaii.edu
}
}
\maketitle

\begin{abstract}
Introductory Computer Science classes are important for laying the foundation for advanced programming courses. However, students without prior programming experience may find these courses challenging, leading to difficulties in understanding concepts and engaging in academic dishonesty such as plagiarism. While there exists plagiarism detection techniques and tools, not all of them are suitable for academic settings, especially in introductory programming courses. This paper introduces PasteTrace, a novel open-source plagiarism detection tool designed specifically for introductory programming courses. Unlike traditional methods, PasteTrace operates within an Integrated Development Environment that tracks the student's coding activities in real-time for evidence of plagiarism. Our evaluation of PasteTrace in two introductory programming courses demonstrates the tool's ability to provide insights into student behavior and detect various forms of plagiarism, outperforming an existing well-established tool.

A video demonstration of PasteTrace and its source code, and case study data are made available at \url{ https://doi.org/10.6084/m9.figshare.27115852}
\end{abstract}


\section{Introduction}
Introductory Computer Science classes serve as the initial exposure to programming for many students, playing a vital role in building the foundation for subsequent higher-level courses \cite{omer2021introductory}. It is through these courses students learn about the fundamentals of programming, such as variables, data types, control structures, and functions, which they build upon in their subsequent higher-level courses \cite{Piwek2020,nikula2007python}. Moreover, the importance of these courses extends beyond students learning about programming language semantics and syntax. Through these courses, students also develop and practice problem-solving, creativity, and logical reasoning skills, which they can apply to their programming assignments \cite{Medeiros2019}.

However, these courses can be challenging for many students, especially those with no prior programming experience. These challenges range from syntactic difficulties to conceptual misunderstandings as well as deficits in problem-solving, code organization, and debugging \cite{Qian2017}. Consequently, students may unintentionally or intentionally engage in academic dishonesty, such as plagiarism \cite{fraser2014collaboration}.

Plagiarism is particularly troublesome in introductory classes.  Since an introductory class is often a student's first experience of programming, if a student passes an introductory class by plagiarizing, then it is likely that the student does not understand part or all of the course. If students are not corrected for plagiarizing in their introductory programming classes, they may develop a false sense of their abilities. This lack of understanding due to plagiarism could leave them unprepared for more advanced courses and future career challenges.

\subsection{Goal}
In the field of computing education, several techniques and tools have been developed to identify different types of plagiarism \cite{martins2014plagiarism, jiffriya2021plagiarism}. While these tools are valuable, not all of them are suitable for academic settings, especially in the context of introductory programming courses. Some of these tools are proprietary and closed-source, making it challenging for institutions to assess their accuracy and reliability. Additionally, certain plagiarism detection tools require the transfer of files to external services, raising concerns about potential violations of student privacy laws.

The goal of our research is to provide an effective and reliable open-source code plagiarism detection tool specifically tailored for academic environments, particularly in introductory programming courses. To this extent, we present \textit{\ToolName, a single source plagiarism detection tool For introductory programming courses}.

\section{Background}
Plagiarism is generally defined as ``The practice of taking someone else's work or ideas and passing them off as one's own.'' \cite{oxed}  However, for this paper, we define plagiarism as ``\textit{copying someone else's code to complete an assignment}.''  We exclude copying ideas for several reasons, mainly because it is difficult to determine if an idea has been copied or just independently rediscovered; the general culture of computer science is to iterate on others ideas (often without attribution \cite{gibson-reuse}); and because at the introductory level, all the required ideas have been thought of by so many people that determining an origin would be next to impossible.  Additionally, we exclude the possibility a student could ``cite'' their copied code to avoid plagiarism because the introductory level involves teaching coding, not copying existing code.

\subsection{Existing Techniques for Detecting Code Plagiarism}
The general method of plagiarism can be summarized as follows:  Given a set $A$ of submissions, find a set $P$ such that  $\forall p\in A,$ if $\exists q\in A$ such that $q\not=p$ and $D(p,q)$, then $p\in P$ and $q\in P$ for some plagiarism detection function $D$.  However, defining $D$ effectively can be challenging \cite{attitude}. 
	
Researchers at Stanford University developed a program called MOSS (Measure of Software Similarity), which is capable of giving each assignment pair a similarity score \cite{moss}. This approach is robust against name changes, code reorganization, and white space changes by analyzing a semi-random subset of k-gram hashes based on each assignment where k has been manually tuned for each language.
	
Others have attempted to define $D$ by ignoring syntax and exclusively using code semantics \cite{semantic} so that two pieces of code match when they functionally do the same thing.  
There are also many other efforts to define $D$ using a machine learning model \cite{uu_thesis,IIITH_thesis,mp_ml,ml_style}. Additionally, there are also detection models that attempt to exclusively detect single LLMs output \cite{llm_detect}. However, these are not particularly reliable at present\cite{23_percent} and have a high false positive rate\cite{and_you_fail}. AlSallal et al. \cite{english} used machine learning to detect stylistic variation in essays. However, in our scenario, this approach would not be ideal for introductory-level students who may not yet have a defined ``style.''

\section{\ToolName} 
\subsection{Single Source Plagiarism Detection}
\ToolName is an Integrated Development Environment (IDE) based on the Processing IDE.  Processing is an open-source IDE for Java that is designed to be easy to learn even for non-programmers \cite{Greenberg2007}; as such, it is ideal for our application.  

When the modified IDE first launches, a persistent UUID file is created on the machine that installed it.  This UUID acts as a machine or user ID, is referred to as the \installID, and is assumed to be constant throughout the class.  Additionally, whenever a project is created with the IDE, a second UUID is created that is unique to that project. We refer to this as the \projectID.  Both UUIDs are saved in a special hidden metadata comment in the program source file. We will refer to this as the \metaComment.

Whenever a file is opened, the IDE checks the \installID of the file and machine.  If they differ, the IDE notes the new \installID in the \metaComment in an ordered list called the \infectionStack.  A similar stack is also used to track paste events.  As these stacks are effectively a single stack with a way to discriminate between event types, we will use \infectionStack to refer to both.

Additionally, when any part of the code is copied, zero-width spaces (U+200B) are interspersed with the normal letters to encode as much data as possible.  The exact amount of data varies based on the size of the copy, but it can include several copies of the \installID, \projectID, and \infectionStack.  When pasted into a project, the encoded data is compared to that of the project receiving the code, and any mismatch UUIDs are added to the \infectionStack. This encoding survives being sent over most messaging software such as Discord, and is generally resilient to partial pastes.  Any paste without encoded data is logged as originating outside the IDE. We should note that our testing revealed that all email clients strip out the zero-width spaces, leading to emailed code copies being identified as originating from outside the IDE. 

In addition to copy tracking, the IDE also logs the time, location, and content of all edit events, including copy, paste, cut, delete, and typing.  Any mismatched UUIDs are included in the log for the respective paste event.  This data is also included in the \metaComment.  Using this keylog, it is possible to reconstruct the entire coding process and trace all code pastes. Additionally, if the IDE loads a file without a \metaComment, it notes this as the first edit and logs the initial state of the file.

Further, the UUIDs can be used to trace code authorship through multiple student submissions.  Additionally, analyzing an entire class at the same time can be useful to identify false positives, such as copies from previous assignments on the student's machine.

Even simple files such as Hello World would be detected, as the detection does not rely on code comparison. Our method is immune to traditional false positives as organically created code is never considered plagiarized regardless of similarity to another student.

\subsection{Plagiarism Detection Scenarios}

\subsubsection{Peer-to-peer (P2P) file sharing.} In this situation, one student writes the code and shares it with a peer, who then submits the copy. With \ToolName, there is no easy way to submit someone else's unmodified file without the UUIDs being a clear red flag, and any attempt to edit the file will mark it as infected. Additionally, introductory classes often require students to add an "Author" comment to the top of their files. This would need to be edited by anyone attempting to cheat, and the original comment is preserved in the log. Finally, any code shared via online messaging will be flagged and, in general, traced. If a student copies from a previous assignment, it will be traceable to which assignment and who authored it.

\subsubsection{Collaboration.} In this situation, two students work together on an assignment that is supposed to be completed individually. Collaborating students will have similar code with overlapping timestamps at each edit.  Additionally, collaborating students tend to send code back and forth, meaning both side's \infectionStack will be populated with each other's UUIDs.  This back and forth should make it easy to distinguish from P2P.
	
\subsubsection{Theft.} This is any form of P2P plagiarism where the author does not know that it is happening.  For example, a student (the author) works on a shared computer and leaves a file saved locally, and the plagiarizer finds the file when attempting the same assignment.  
While it can be challenging to discriminate it from P2P, theoretically, it is possible to isolate the original file using the knowledge of the computer setup (e.g., a computer lab with known \installID{}s), the \infectionStack, and the timestamp log.  

\subsubsection{Search.} This is when a student uses an internet search engine to find an existing solution to an assignment. No matter how a file is found, once it is loaded by the IDE, it will be marked as an external file in the \metaComment. If it is submitted without being opened by the IDE, it will lack a \metaComment.  If the source code is copied into an existing project, it will be marked as external. The only viable source that this does not detect is physically typing out the code from a reference.  However, this is theoretically detected by analyzing the activity.

\subsubsection{Expert.} Similar to the Search scenario, the Expert scenario involves obtaining a solution from paid sources (e.g., Chegg or Coursehero), or the solution is tailored by an expert to the exact assignment given. Expert sources are detected in the same way as Search.

\subsection{Automated Detection}
To facilitate a batch evaluation of student submissions, we implemented a script that takes all student submissions for a given assignment and analyzes them.
The script is designed to analyze a directory of student submissions by building a graph using the \infectionStack to trace shared files and identify shared machines. It categorizes paste events and detects plagiarism based on various criteria, such as the origin of pastes and the number of lines pasted. Purely internal paste events, or ones that are too short to track, are ignored.  From there, pastes fall into 4 categories:  pastes originating from a project that was turned in by another student, or pastes originating from a different students' machine; pastes originating from the same machine, but a different project; pastes that originate from an \installID that is not associated with any project turned in; and pastes of external origin.  In each case where plagiarism is detected, the code pasted is also logged for human validation.

Pastes from another turned-in \projectID or \installID that is associated with another student are immediately marked as plagiarism. Pastes that originate on the same machine are treated differently depending on the machine. If the machine was not shared, the paste is assumed to originate from the same student. However, there is the possibility they are trying to use a loophole where the assignment is completed in another project and then pasted to the final project to clean up the keystroke log. To detect this, we decided to use a 50-line threshold. This allows functions to be copied from previous assignments but still detects entire projects being copied. If the machine was shared, the paste is flagged as plagiarism.

Pastes that originate from a different machine that is not associated with any student and a project not associated with any student are assumed to be from the same student, similar to pastes from the same machine; however, as this is less likely to be the student and may be an expert source, we have set the threshold to 20 lines. 

Pastes that did not originate in the IDE are treated as copying from online resources.  While a certain amount of online copying is permissible (like copying the name of a function from a docs page), copying whole functions would not be.  We set the threshold for these events at three lines, reasoning that 1 line is likely a line from a docs page, and three lines are likely accidentally copying a new line before and after the function within the docs page, but more than three lines could be an entire function.

Additionally, we track the number of edits made after the last plagiaristic paste event.  

\section{\ToolName Evaluation}
\label{sec:evaluation}
To assess the effectiveness of \ToolName, we conducted a human-subject based study with students enrolled in Introduction to Programming I and Introduction to Programming II courses at the University of Hawai`i at M\=anoa. We collaborated with the instructors to provide each class with a project that was challenging enough to encourage plagiarism, but not so difficult that no students could complete it on their own. One class was tasked with creating three polymorphic-shaped objects that move around the screen and bounce off the edges. The other class was given the task of creating a function-defined fractal, such as a Mandelbrot fractal. Students were instructed to use \ToolName and were expressly allowed to plagiarize the assignment. Extra credit was offered to students who participated in this activity.  Further, we seeded two copies of the full solution on the internet for each assignment.

The participating students were provided with a USB flash drive containing \ToolName, which was configured to save project files within the flash drive. Participants received extra credit when they turned in the flash drive. Further, students were asked to self-report if they plagiarized to complete the assignment. 
We should note that the data collection was anonymous; the flash drive and its contents, along with self-reported data, cannot be tied to any specific student. 

As our evaluation was conducted by students enrolled in two different courses, we report on our findings for each class separately. As part of our evaluation, we also analyzed the submitted code using a state-of-the-art tool, i.e., MOSS. The files given to MOSS did not include our \metaComment.

MOSS takes all code files submitted for an assignment as input and then returns a list of file pairs.  Each pair is given a percentage score that represents what percentage of the firs file is also in the other file.  Pairings are not necessarily bijective, and there can be up to 2 items per pair of files, each with a different percentage depending on which file is listed first.  The MOSS developers refer to this as the "Measure of Software Similarity" and emphasize that similar is not necessarily plagiarized, and manual review is required for each pair.  MOSS helpfully highlights the shared code.  Only file pairs with some similarity are returned.  We manually inspected all detected pairings for the 'MOSS Output' Column of Tables \ref{tab:ics111} and \ref{tab:ics211}.  As MOSS results require interpretation, we categorized MOSS results in 3 ways.
\begin{enumerate}
    \item Major Link: A large portion of the code is shared with the listed file, enough that we would have counted the file as plagiarized had this been a regular assignment.  While this is an arbitrary line, for these assignments "large" is "more than 40\% lines similar."
    \item Minor Link: Moss reported that there is a link, but after review, we determined this link was not significant and was only composed of code that would naturally be similar given the problem (e.g. a double nested for-loop using x and y when iterating over an image).  While again arbitrary, for these assignments "less than 40\%"  was the dividing line.
    \item No Plagiarism: Moss reported no links where this file was the first in the pair.
\end{enumerate}
Note that in these case studies, there is a clean divide at 40\% between Plagiarized, and not, but when using MOSS in the real world, there is often not a clean division, and the pairs require more thorough inspection to determine severity. 

The code, raw and pre-processed data, automation reports, and MOSS reports are available online\footnote{https://figshare.com/s/43fbf89a815ba5c248cb}.

\subsection{Evaluation Results: Introduction to Programming I}
    \label{sec:casestudy111}

\begin{table*}[thb]
    \centering
    \caption{Summary of evaluation results for Introduction to Programming I}
    \label{tab:ics111}
    % {\large{\phantom{A}}} makes rows taller
    \begin{tabular}{|l|l|l|l|l|}
        \hline
        Student & Method & Visible in \metaComment & Automated Check output & Moss Output \\
        \hline
        A & Plagiarized & Records large paste & Plagiarism Detected, 35 Edits & No Plagiarism \\\rowcolor{Gray}
        B & Legitimate & No Warning Signs & No Plagiarism Detected & No Plagiarism \\
        C & Unclear & Large internal paste & Likely Plagiarized, 31 Edits & Minor link to E \\\rowcolor{Gray}
        D & Plagiarized& Records large paste & Plagiarism Detected, 31 Edits & No Plagiarism \\
        E & Legitimate & Large Paste followed by many edits & Plagiarism Detected, 767 Edits & Minor link to C \\\rowcolor{Gray}
        F & Plagiarized & Large external paste & Plagiarism Detected, 0 Edits & No Plagiarism \\
        \hline
    \end{tabular}
\end{table*}

We received six submissions from the first course. Each submission has been arbitrarily assigned a single-letter designation from A to F. The results are summarized in Table \ref{tab:ics111}. Out of these six submissions, four successfully completed the assignment, while two (A and D) did not.

First, examining the code submitted by students who did not complete the assignment, we observe that Student A copied and pasted from ChatGPT but prompted it wrong and did a similar but different assignment.  Plagiarism is easily detected as there the \metaComment is just a 75-line paste followed by about 38 minor tweaks and formatting changes. Next, Student D appears to have struggled: the code they copied is a badly mangled AWT example and does not compile without significant modification.

Examining the correct assignments, Student B completed the assignment legitimately, and it is shown in the \metaComment{}s.  There are multiple well-organized files, each organically coded with no large external pastes or irregularly linear typing sections. There are copies between files where code should be reused and modified, as well as follow-up edits to make these modifications. Student E pasted a large section of code from a different project with a matching \installID and then spent considerable time (around 8 hours spread over a week or so) debugging and improving it.  We believe this to be a legitimate completion. Student F found and submitted one of the seeded assignment files.  It is logged as a large foreign paste. Finally, we are unable to determine if the code from Student C is legitimate.

MOSS only identified a small connection between Student C and E.  This connection is entirely composed of the expected boilerplate code.

\subsection{Evaluation Results: Introduction to Programming II}

    \label{sec:casestudy211}
\begin{table*}[thb]
    \centering
    \caption{Summary of evaluation results for Introduction to Programming II}
    \label{tab:ics211}
    % {\large{\phantom{A}}} makes rows taller
    \begin{tabular}{|l|l|l|l|l|}
        \hline
        Student & Method & Visible in \metaComment & Automated Check output & Moss Output \\
        \hline
		A & Plagiarized&Records multiple large pastes & Plagiarism Detected, 1 Edits & No Plagiarism \\\rowcolor{Gray}
		B & Plagiarized&Records multiple large pastes & Plagiarism Detected, 0 Edits& Minor link to R \\
		C & Plagiarized& Records multiple large pastes &Plagiarism Detected, 4 Edits & Minor link to R \\\rowcolor{Gray}
		D & Plagiarized&Records large paste & Plagiarism Detected, 0 Edits & Minor link to Q \\
		E & Plagiarized& Visible Linear Coding & No Plagiarism Detected &  Major Link to F, I, H, and P\\\rowcolor{Gray}
		F & Plagiarized&Records large paste &Plagiarism Detected, 105 Edits & Major Link to E, I, H, and P\\
		G & Plagiarized&Records large paste &Plagiarism Detected, 78 Edits & No Plagiarism \\\rowcolor{Gray}
		H & Plagiarized&Records large paste &Plagiarism Detected, 78 Edits& Major Link to E, F, I, P \\
		I & Plagiarized&Infected by H \installID &Copied Directly from H& Major Link to E, F, H, P  \\\rowcolor{Gray}
		J & Blank & Blank Submision & &   \\
		K & Blank & Blank Submision & &\\\rowcolor{Gray}
		L & Plagiarized&Records multiple large pastes  &Plagiarism Detected, 11 Edits &No Plagiarism \\
		M & Unclear& Records Large Internal Paste & Plagiarism Detected, 0 Edits&No Plagiarism \\\rowcolor{Gray}
		N & Legitimate&No Warning Signs & No Plagiarism Detected & No Plagiarism\\
		O & Plagiarized &Records multiple large pastes  &Plagiarism Detected, 303 Edits & No Plagiarism \\\rowcolor{Gray}
		P & Plagiarized&Records large paste &Plagiarism Detected, 38 Edits& Major Link to E, F, I, H \\
		Q & Plagiarized&Records large paste &Plagiarism Detected, 3 Edits & Minor link to D and R\\\rowcolor{Gray}
		R & Legitimate&No Warning Sign & No Plagiarism Detected& Minor Link to Q\\   
        		
        \hline
    \end{tabular}
\end{table*}
We received 18 submissions from the second course. Each submission has been given a single-letter designation from A to R. The results are summarized in Table \ref{tab:ics211}. Out of these 18 submissions, 11 successfully completed the assignment, while 5 (C, G, L, M, and N) did not, and 2 (J and K) were empty flash drives.

First, we examine the incorrect submissions. Student C submitted a working Mandelbrot fractal written using Swing, but it does not run in our IDE. Student G submitted copied code for a Koch Snowflake written in Processing, which also does not run. Student L initially copied the code of a Swing program that makes a 3D rendering of a rain of rotating toroids and then converted it to run in Processing. However, this is not a solution to the actual assignment given. Student M copy and pasted their code from another project on their computer, thus destroying the typing event history. They claimed an IDE bug was preventing their code from saving. In any case the solution provided did not meet the assignment's requirements. Student N has created a 3D solar system simulation using rotation and translation matrices. However, it does not meet the assignment requirements.

We are unsure how C and G managed to find broken solutions before finding working solutions. We tried searching for the original source code, but were unable to locate it.  During the search we did find many solutions that would have worked, but apparently the students did not find these.

Next, we examine the correct assignments. From the 11 submissions, 5 (E, F, H, I, and P) were flagged by MOSS as having quite similar solutions with a similarity score of about 50\%, so we shall discuss those first.

Student E's solution contains no large external pastes. However when the edits are animated, the result is visibly linear coding as if typing.  The student claims they watched a YouTube video of someone coding a Mandelbrot fractal and followed along. 
While we consider E's solution to be plagiarized, it is not part of the larger group unless the group copied a source that itself is linked to the Video that E copied. Student F and P independently copied a stock example from the Processing website\footnote{https://processing.org/examples/mandelbrot.html}. Student H followed an online video.
After which, the final code was sent to I.  The \metaComment for I is corrupted after this paste, but the final solution of both students was significantly modified after from this paste event, and the paste has little resemblance to either final submission.  The \infectionStack successfully tracks this exchange.

Student A copied a Penrose Snowflake example using multiple copies with no meaningful modifications. Likewise, student B copied a Mandelbrot Fractal and made no meaningful modifications. Students D and Q both asked ChatGPT to generate a Mandelbrot and then copied the result without modification.  The results are nevertheless distinct from each other. Student O copied a tree fractal example and then modified it to be slightly different. Student R claims to have used a video to help them make a Mendelbrot fractal complete with zoom. However, there is no indication they copied from the video, and we believe this to be a legitimate completion of the assignment.

MOSS revealed weak connections between B and R, and C and R, which were deemed to be incidental. Additionally, a moderate connection was identified between D and Q, as well as R and Q, despite the absence of a direct connection between D and R. Notably, only two of these connections were generated by the ChatGPT platform. Furthermore, the identification of a strong connection between H and I, and between F and P, was indicative of plagiarism within those respective pairs. However, MOSS also flagged weaker yet still notable connections among all four pairs and additionally with E.  It may not be accurate to include E in this group, as E used a different source. It can be argued that MOSS correctly identified the plagiarism despite the changes. However, the similarities between the two groups are the same as the similarities between each and E, and the code is quite common to Mandelbrot fractals.  As such, we believe this to be incorrect with regard to plagiarism.

\subsection{\ToolName Vs. MOSS}
Comparing directly to MOSS objectively is difficult as MOSS requires a highly subjective analysis.  However, if each reported pairing from MOSS is considered correct, then MOSS successfully detected 9 of 16 plagiarised assignments and had two false positives. However, if our subjective analysis is used, and only Major links are considered, it falls to 5 of 16 with 0 false positives.   In comparison, our tool detected 15 of 16 and had one false positive.   

Due to its inherent design, MOSS is incapable of detecting plagiarism unless two students have submitted similar code files.  Since our tool relies on the method used to create the code, not just the end result, it can detect such events.

\section{Conclusion}
This paper introduces PasteTrace, a novel open-source plagiarism detection tool specifically designed for introductory programming courses. Our evaluation of PasteTrace demonstrates the tool's capability to detect various forms of plagiarism, including those from single sources, which are often overlooked by comparative tools like MOSS. Additionally, the tool's ability to track copy-paste events, code origins, and the overall coding process provides valuable insights into student behavior and potential academic dishonesty. Our future work for PasteTrace includes refining the detection algorithms, expanding compatibility with various IDEs, and conducting large-scale studies across diverse programming courses.

\bibliographystyle{ieeetr}
\bibliography{main}
\end{document}